\documentclass[envcountsame,runningheads,notitlepage]{llncs}

\usepackage{listings}
\usepackage{color}
\usepackage{amsmath}  
\usepackage{graphicx} 
\usepackage{amssymb} 
\usepackage{algpseudocode}
\usepackage{algorithm}
\usepackage{comment}
\usepackage{hyperref}
\usepackage{orcidlink}

\usepackage{float}
\floatstyle{plaintop}
\restylefloat{table}

\title{Breaking ECDSA with Two Affinely Related Nonces}
\titlerunning{Breaking ECDSA with Two Affinely Related Nonces}

\author{
  Jamie Gilchrist\orcidlink{0009-0008-0649-4263}\inst{1} \and
  William J. Buchanan\orcidlink{0000-0003-0809-3523}\inst{1} \and
  Keir Finlow-Bates\orcidlink{0009-0004-8308-615X}\inst{2}
}

\institute{
  \inst{1} Blockpass ID Lab, Edinburgh Napier University, UK \and 
  \inst{2} Chainfrog Oy., Eura, Finland
}

\begin{document}
\maketitle
\begin{abstract} 
The security of the Elliptic Curve Digital Signature Algorithm (ECDSA) depends on the uniqueness and secrecy of the nonce, which is used in each signature. While it is well understood that nonce $k$ reuse across two distinct messages can leak the private key, we show that even if a distinct value is used for $k_2$, where an affine relationship exists in the form of: \(k_m = a \cdot k_n + b\), we can also recover the private key. Our method requires only two signatures (even over the same message) and relies purely on algebra, with no need for lattice reduction or brute-force search(if the relationship, or offset, is known). To our knowledge, this is the first closed-form derivation of the ECDSA private key from only two signatures over the same message, under a known affine relationship between nonces.

\end{abstract}


\section{Introduction}
The Elliptic Curve Digital Signature Algorithm (ECDSA) method has been around for over two decades and was first proposed in \cite{johnson2001elliptic}. It is a widely used cryptographic method for generating digital signatures. It plays a crucial role in ensuring the authenticity, integrity, and non-repudiation of digital messages or transactions. Based on elliptic curve cryptography (ECC), ECDSA offers strong security with relatively small key sizes, making it more efficient than traditional algorithms like RSA in terms of speed and resource usage.

ECDSA is popularly used in a variety of modern digital systems where security and efficiency are critical. In the world of cryptocurrencies, it is used to secure Bitcoin and Ethereum transactions by allowing users to sign transactions with their private keys, proving ownership without revealing sensitive information. ECDSA is also used in secure communication protocols such as TLS (Transport Layer Security), which underpins HTTPS and ensures safe web browsing by verifying the authenticity of websites. Additionally, it is employed in secure email systems through standards like S/MIME, passkey authentication protocols such as FIDO2, and in mobile authentication, particularly within modern smartphones and IoT devices, where its lightweight nature is ideal for constrained environments.

ECDSA was standardised by the National Institute of Standards and Technology (NIST) in 2000 and has since remained a widely used digital signature scheme. Alongside, RSA (Rivest, Shamir, Adleman) and EdDSA (Edwards-Curve Digital Signature Algorithm), ECDSA is standardised in FIPS 186-5 \cite{nistFederalInformation}. However, it poses several challenges for inexperienced implementers, most notably the risk of nonce reuse attacks, which can compromise the security of the private key if not properly mitigated.

In this paper, we begin with an overview of the mathematics behind ECC and ECDSA. We then proceed to outline a novel method for private key recovery, where only two signatures are needed(even over the same message) when we know there was an affine relationship between any two or more nonces used during independent signatures.

\section{Background}
This section outlines some of the core theory related to elliptic curves. 
\subsection{Finite integer fields and modular arithmetic}
Elliptic curve cryptography takes place over a finite integer field (also called a Galois field) of prime order. Given a prime number $p$, we take the set of all integers from 0 to $p-1$, and use addition and multiplication modulo $p$ for our calculations. In mathematics, this is denoted by $\mathbb{F}_p$.

Modular multiplication and addition are associative, distributive, and commutative. This is extremely helpful for developers because it means that, provided your programming language implements arithmetic in a manner that supports large enough numbers, you can apply the modulo operations at any convenient point (provided you also remember to apply them at the end of your calculations).

\subsubsection{Associativity}
\begin{align*}
(a \bmod p + b \bmod p) \bmod p = (a + b) \bmod p\\
(a \bmod p \cdot b \bmod p) \bmod p = (a \cdot b) \bmod p
\end{align*}

\subsubsection{Commutativity}
\begin{align*}
a \bmod p + b \bmod p = b \bmod p + a \bmod p\\
a \bmod p \cdot b \bmod p = b \bmod p \cdot a \bmod p
\end{align*}

\subsubsection{Distributivity}
\begin{align*}
(a \bmod p + b \bmod p) \cdot c \bmod p = (a \cdot c + b  \cdot c) \bmod p\\
c \bmod p \cdot (a \bmod p + b \bmod p) = (c \cdot a + c \cdot b) \bmod p
\end{align*}

In the text that follows, it should be understood that the numbers we are working with are all integers in the range $\{0, ... , p-1\}$, and all operations are conducted modulo the order of the finite field over which mathematical operations take place.

\subsection{Elliptic curves}

Elliptic curves are the subject of a wide and complicated branch of mathematics, finding uses in all sorts of unexpected areas, such as Andrew Wiles's proof of Fermat's Last Theorem and, importantly, cryptography.  The Elliptic Curve's application to cryptography was first proposed in the mid-1980s by Neal Koblitz and Victor Miller \cite{koblitz1987elliptic} \cite{miller1986use}

Fortunately, implementing ECC doesn't require a developer to select optimal curves and parameters from scratch themselves. Standards have emerged which recognise certain elliptic curves with carefully vetted cryptographic properties, which are then used in publicly available software libraries. A list of well-tested elliptic curves and their commonly used parameters can be found at \cite{sec2v2}. For example, Bitcoin and Ethereum use an elliptic curve identified as $secp256k1$, and FIDO2 uses $secp256r1$.

An elliptic curve is the set of solutions $(x, y)$ to an equation of the form:
\begin{align}
y^2 = x^3 + ax + b
\end{align}

For ECC, we require a definition of scalar multiplication of a point on the curve. 


Given two points $P$ and $Q$ on the curve, adding $P$ to $Q$ produces a new point $R$, also on the curve, using the following definition:

\vspace{1em}

Let $P = (x_1, y_1)$, $Q = (x_2, y_2)$, and $R = P \oplus Q = (x_3, y_3)$

\begin{center}
\renewcommand{\arraystretch}{1.4}
\begin{tabular}{@{} l l @{}}
\textbf{Operation}      & \textbf{Formula (mod $p$)} \\[0.5ex]
$P \neq Q$           & $m = \frac{y_2 - y_1}{x_2 - x_1}$ \\
$P = Q$              & $m = \frac{3x_1^2 + a}{2y_1}$ \\
Resulting Point          & $x_3 = m^2 - x_1 - x_2$, $y_3 = m(x_1 - x_3) - y_1$ \\
\end{tabular}
\end{center}

This allows us to define the scalar multiplication $\times$ of a point $P$ by a scalar value $n$:

\begin{align}
n \cdot P = \underbrace{P \oplus P \oplus \cdots \oplus P}_{n\ \text{times}}
\end{align}

In cryptographic libraries, scalar multiplication of curve points is implemented using techniques to improve performance, such as with the Montgomery Ladder method \cite{montgomery1987speeding}. We need to make sure the multiplication is performed securely and constantly timed to prevent side-channel attacks that could reveal information about the values of the operation, such as for constant time computation \cite{bernstein2017montgomery}. As we will be using private keys in the scalar multiplication of a specific curve point called the \emph{Generator Point}, this is important.

The Generator Point $G$ is a point on the elliptic curve (an \(x\) and \(y\) coordinate) which is chosen as the starting point for the point addition. This Generator Point is consistent for any given Elliptic Curve; that is, anyone using the same curve(such as secp256k1, secp256r1, and so on) will also use the same Generator point, as these are part of the standard.


Scalar multiplication on elliptic curves is what provides the cryptographic hardness behind ECC. Given a private key scalar \(priv\) and the generator point \(G\), we can compute the public key as:
\begin{align}
pub = priv \cdot G 
\end{align}
This operation is straightforward to perform using repeated point addition, as we described earlier. However, if we wanted to reverse this(to go back to the private key), that is, given \(G\) and \(pub\), attempting to recover \(priv\) is known as the \textit{Elliptic Curve Discrete Logarithm Problem} (ECDLP), which is considered computationally infeasible for sufficiently large fields and secure curves.

Unlike multiplication over integers, scalar multiplication on an elliptic curve does not have a simple inverse operation. There is no efficient algorithm known that can derive \(priv\) from \(pub = priv \cdot G\) without performing a brute-force search. This "one-way" nature is what makes ECC secure because you can safely share your public key without revealing your private key, even though they are mathematically linked.

\subsection{Digital signing}
To produce an ECDSA signature, we sign a message $M$ using a private key $priv$ and a randomly generated value $k$, called a nonce. The value $priv$ is randomly selected from the underlying set of $\mathbb{F}_p$ and kept secret, but is reused. For every signature, a new value of $k$ is also randomly selected from the set but is never reused, for reasons that will become apparent.

We prove the signature with the public key $pub$ derived from the private key $priv$ using the one-way function shown below.


\begin{align}
\label{pub}
pub = priv \times G
\end{align}

Thus, the public key is a pair of numbers, as it is a point on the elliptic curve. As the elliptic curve is symmetric about the $x$-axis, to store the public key value, we only need to store the $x$ value of the key and the sign of the $y$ value. We can then compute the $y$ value from the elliptic curve equation using $x$ and the sign when needed.

Equation \ref{pub} is a one-way function, also called a trapdoor function. It turns out that calculating $pub$ from $priv$ and $G$ is relatively simple, but reversing the function and deriving $priv$ from $pub$ is infeasible with currently known techniques. This is what allows us to keep the private key secret while revealing the public key.

Each time we sign a new message, a new random nonce value must be used, producing a different (but verifiable) signature. Overall, the signer only has to reveal the elements of the signature and their public key, and not the nonce value.

An ECDSA signature consists of a pair of numbers, $(r, s)$, which are produced as follows:

\begin{align}
r = k \cdot G\\
s = k^{-1} (H(M) + r \cdot priv)
\end{align}

The value $r$ is the $x$-coordinate of the point $k \cdot G$, and $H(M)$ is the SHA-256 hash of the message ($M$) converted into an integer value.

\section{Related work}


Many of the pitfalls that exist in ECDSA due to improper nonce handling have been well documented. \cite {buchanan2025ecdsa} provides a comprehensive breakdown of the most significant attack vectors. With the \textbf{revealed nonce attack}, an exposure of a single nonce can lead to direct private key recovery \cite{asecuritysite_66284}. If the signer accidentally exposes even a single nonce, an attacker can then directly compute the private key using algebra. For the \textbf{fault attack}  we only require two signatures, and where one is produced without a fault $(r,s)$, and the other has a fault $(r_f,s_f)$. From these, we can generate the private key \cite{sullivan2022open,poddebniak2018attacking}. For a \textbf{Weak nonce attack} we can simplify the computation to a discrete logarithm problem with the Lenstra–Lenstra–Lovász (LLL) method \cite{lenstra1982factoring} \cite{asecuritysite_79512}. For the \textbf{nonce reuse attack},  it is well-known that simply keeping the selected nonce secret is not enough to secure the private key \cite{brengel2018identifying}. If a nonce is used to sign a first message to produce a first signature ($s_{1}$) and is then reused to sign a second message to produce a second signature ($s_{2}$), then $s_{1}$ and $s_{2}$ will have the same $r$ value, and it is possible to derive the private key ($priv$) from the two signatures.

A related research effort is presented by Macchetti \cite{cryptoeprint:2023/305}, which explores the case of related nonces, through an unknown algebraic recurrence, potentially of high degree, such as for linear congruential, quadratic or cubic. The method used by Macchetti expresses each nonce as a function of the message hash and signature components, and then recursively eliminates unknown coefficients. The result is a univariate polynomial of the private key, where the private key is then recovered as a root of this polynomial. This technique is powerful(and general), but requires multiple distinct signatures(typically 4 or more), and symbolic solving over finite fields.

In contrast, the method we present focuses on the specific case where two nonces are affinely related with \textit{known} coefficients, a scenario found in flawed implementations using counters, or other linearly based PRNGS. Under this assumption, we show that the private key can be derived directly(in closed form) using only two signatures and pure algebraic manipulation. In our method, there is no need for symbolic tools, lattice reduction, or brute force search(if the relation is known). 

Further, Macchetti's technique assumes that all signatures are over distinct messages. If the message hashes are identical across signatures, the resulting expressions will become structurally dependent, causing the recurrence resolution to fail. Our approach, however, remains valid even when the same message is signed multiple times, with affinely related nonces. This allows our attack to apply to a broader set of flawed systems, which include those where repeated signing of identical payloads occurs.

The focus of this paper is to explore a lesser-known misuse of the nonce value in ECDSA, a \textbf{Linear relationships between nonces}

\section{Affine relationships between nonces}
What is less well-known is that even if two distinct values for $k$ are used for producing two different signatures(even over the same message) with the same private key, if there is a \textit{known affine relationship} between the two values, then the private key can also be extracted.

Consider the situation where Bob generates an initial random value for $k$, and subsequent values are produced using a linear equation of the form:

\begin{align}
k_{n+1} = ak_n + b
\end{align}

For example, if $a=1$ and $b=1$, Bob is using a simple counter for generating values of $k$. Even if the initial nonce is selected randomly, knowing the relationship, we can retrieve the private key.

Once again, we start with two hashes of two messages, $h_1$ and $h_2$, and two nonces, $k_1$ and $k_2$ (where $k_2 = ak_1 + b$), resulting in two signatures, $(r_1, s_1)$ and $(r_2, s_2)$. Using the ECDSA signature equation:
\begin{align}
s = k ^{-1} (h + r \cdot priv) \mod n
\end{align}

where we can then rearrange it to express $k$ in terms of known quantities:

\begin{align}
    k_1 &=\frac{h_1 + r_1 \cdot priv} {s_1} \label{eq:k1} \\
    k_2 &= \frac{h_2 + r_2 \cdot priv}{s_2} \label{eq:k2}
\end{align}

We then substitute the affine relation $k_2 = a k_1 + b$ into equation \eqref{eq:k2} :
\begin{align}
\frac{h_2 + r_2 \cdot priv}{s_2} = a \cdot \frac{h_1 + r_1 \cdot priv}{s_1} + b
\end{align}

Multiply both sides by $s_2$, then expand:
\begin{align}
h_2 + r_2 \cdot priv = \frac{a s_2}{s_1}(h_1 + r_1 \cdot priv) + b s_2
\end{align}
Move all $priv$ terms to one side:
\begin{align}
r_2 \cdot priv - \frac{a s_2 r_1 \cdot priv}{s_1} = \frac{a s_2 h_1}{s_1} + b s_2 - h_2
\end{align}

Factor out $priv$
\begin{align}
priv \cdot \left( r_2 - \frac{a s_2 r_1}{s_1} \right) = \frac{a s_2 h_1}{s_1} + b s_2 - h_2
\end{align}

Finally, we can solve for the private key:

\begin{align}
\label{eq:privkey_linear}
priv = \frac{a s_2 h_1 - h_2 s_1 + b s_1 s_2}{r_2 s_1 - a r_1 s_2} \mod n 
\end{align}

The equation allows for the recovery of the private key using only two signatures, the message hashes and knowledge of the affine relationship of the nonces used during signing. Importantly, this works even if the messages are identical.

\subsection{Brute-forcing nonce relationships}
Although software implementations of ECDSA should be made open source to allow third parties to detect vulnerabilities, in practice, it is not possible to determine which software package or application was used to generate signatures when viewing them raw. For example, the Bitcoin blockchain is full of ECDSA signatures, but we have no way of knowing which blockchain wallet(vendor or software version) was used for signing.

In the case where a known relationship between nonces is not explicitly known, an attacker can attempt a brute-force approach by iterating through candidate values for \(a\) and \(b\), testing whether the resulting private key is valid. This technique is computationally more expensive but still feasible when the relationship uses small constants or predictable patterns.

%

\section{Conclusionx}
The main focus of this paper has been to outline how an affine relationship between nonces used in two ECDSA signatures can be exploited to derive the private key. However, we wonder whether this can be extended further, where we can solve for relationships that can be considered quadratic or in higher order. (e.g., \(k_2 = k_1^2 + c\)). We believe that this can form the basis of a further research paper.

This paper has demonstrated that the security assumptions of ECDSA collapse whenever two distinct nonces are linked by a known affine relation $k_2 = a k_1 + b$.  Unlike previous related nonce attacks which require four or more signatures, our derivation shows that only two signatures, even over the same message, are sufficient to recover the private key in closed form.  The attack relies on nothing more than modular arithmetic and modular inversion, making it suitable for deployment, once the coefficients $a$ and $b$ are known(or guessed)

Although we focused on first‑degree affine relations, the algebraic approach invites several possible extensions: Exploring quadratic or higher‑order correlations, partial‑information scenarios where only one coefficient is known, and hybrid lattice‑plus‑algebra methods all constitute potential future research. This work reinforces the need for proper nonce generation in ECDSA (either using CSPRNG or with RFC 6979).

\section{Appendix}

\subsection{Exploiting a Known Affine Relationship}

In this section, we will demonstrate how an adversary can recover the private key when the nonces used across two signatures are linearly related as:

\begin{align}
k_2 = a \cdot k_1 + b,
\end{align}

and both signatures \((r_1, s_1)\) and \((r_2, s_2)\) are available (either for the same message or different messages), it is possible to derive the private key using only algebra.

This situation may occur when a flawed implementation uses predictable nonces, such as those generated via a counter or a simple linear recurrence, even if the original(initial) nonce was random. 

The following code samples illustrate both the generation of such signatures and the recovery of the private key under the known affine relationship.

\paragraph{Key Generation — \texttt{gen\_keys.py}}

\begin{lstlisting}[language=Python, caption={ECDSA key generation using SECP256k1}, label={lst:gen_keys}]
from ecdsa import SigningKey, SECP256k1

sk = SigningKey.generate(curve=SECP256k1)
vk = sk.verifying_key

priv = sk.privkey.secret_multiplier

# 33 byte public key
x = vk.pubkey.point.x()
prefix = b'\x02' if vk.pubkey.point.y() % 2 == 0 else b'\x03'
compressed = prefix + x.to_bytes(32, 'big')

print(f"PRIVATE_KEY: {priv}")
print(f"PUBLIC_KEY_COMPRESSED: {compressed.hex()}")
\end{lstlisting}

\paragraph{Signature Generation with Affine Nonce — \texttt{sign\_with\_offset.py}}
\begin{lstlisting}[language=Python, caption={Generating two signatures with affinely related nonces}, label={lst:sign_with_offset}]
import sys, hashlib
from ecdsa import SECP256k1


# CLI: python sign_with_offset.py <priv> <a> <b>
priv = int(sys.argv[1])
a = int(sys.argv[2])
b = int(sys.argv[3])

# Base nonce k1
k1 = 34346754854893457289357283057230582930523052375835723057
k2 = (a * k1 + b)  # General affine relation

curve = SECP256k1
G = curve.generator
n = curve.order

# Messages (can be same or different)
m1 = b"Affinely related nonces are insecure"
m2 = b"Affinely related nonces are insecure"

# Hash the messages
z1 = int.from_bytes(hashlib.sha256(m1).digest(), 'big') % n
z2 = int.from_bytes(hashlib.sha256(m2).digest(), 'big') % n

# r values (x-coordinate of k*G)
r1 = (k1 * G).x() % n
r2 = (k2 * G).x() % n

# Signature components
s1 = (pow(k1, -1, n) * (z1 + r1 * priv)) % n
s2 = (pow(k2, -1, n) * (z2 + r2 * priv)) % n

# Output values to feed into recover_key.py
print(f"z1={z1}")
print(f"r1={r1}")
print(f"s1={s1}")
print(f"z2={z2}")
print(f"r2={r2}")
print(f"s2={s2}")
print(f"a={a}")
print(f"b={b}")

\end{lstlisting}

\paragraph{Private Key Recovery — \texttt{recover\_key.py}}

\begin{lstlisting}[language=Python, caption={Recovering the private key from known affine nonce relationship}, label={lst:recover_key}]
import sys

# CLI: python recover_key.py <z1> <r1> <s1> <z2> <r2> <s2> <a> <b>
z1 = int(sys.argv[1])
r1 = int(sys.argv[2])
s1 = int(sys.argv[3])
z2 = int(sys.argv[4])
r2 = int(sys.argv[5])
s2 = int(sys.argv[6])
a = int(sys.argv[7])
b = int(sys.argv[8])

# Curve order for secp256k1
n = 0xFFFFFFFFFFFFFFFFFFFFFFFFFFFFFFFEBAAEDCE6AF48A03BBFD25E8CD0364141

# Equation (7)
numerator = (a * s2 * z1 - s1 * z2 + b * s1 * s2) % n
denominator = (r2 * s1 - a * r1 * s2) % n
priv = (pow(denominator, -1, n) * numerator) % n

print(f"[+] Recovered private key: {priv}")

\end{lstlisting}

\bibliographystyle{IEEEtran}
\bibliography{main}

\end{document}